\documentclass[seceq,supplement]{ptptex}

\usepackage[dvips]{graphicx}
\usepackage[dvips]{color}
\usepackage{amsmath}
\usepackage{amssymb}
\usepackage{bm}
\usepackage{times}
\usepackage{revsymb}
\allowdisplaybreaks

\newcommand{\vct}[1]{\mathbf{#1}}

\newcommand{\kap}{\mbox{\boldmath $\kappa$}}

\newcommand{\Om}{\Omega^\dagger}

\title{
Non-Equilibrium relation between mobility and diffusivity of  
interacting Brownian particles under shear
}

\author{
Matthias \textsc{Kr\"uger} and Matthias \textsc{Fuchs}%
}

\inst{
Fachbereich Physik, Universit\"at Konstanz, 78457 Konstanz, Germany
}

\abst{
We investigate the relation between mobility and diffusivity for Brownian particles under steady shear near the glass transition, using mode coupling approximations. For the two directions perpendicular to the shear direction, the particle motion is diffusive at long times and the mobility reaches a finite constant. Nevertheless, the Einstein relation holds only for the short-time in-cage motion and is violated for long times. In order to get the relation between diffusivity and mobility, we perform the limit of small wavevector for the relations derived previously [Phys. Rev. Lett. {\bf 102} (2009), 135701], without further approximation. We find good agreement to simulation results. Furthermore, we split the extra term in the mobility in an exact way into three terms. Two of them are expressed in terms of mean squared displacements. The third is given in terms of the (less handy) force-force correlation function.
}
\begin{document}
\maketitle
\section{Introduction}\label{sec:FDTFDR}
In thermal equilibrium, the Einstein relation\cite{Einstein05} for a Brownian particle (a colloid) is the most known application of the fluctuation dissipation theorem (FDT). It connects the equilibrium mobility $\mu^{(e)}(t)$ of the particle to its equilibrium diffusivity $D^{(e)}(t)$,
\begin{equation}
\frac{\partial}{\partial t}D^{(e)}(t)\,t={k_BT}\mu^{(e)}(t),\label{eq:Einsteineq}
\end{equation}
with  thermal energy $k_BT$.
The diffusivity on the left hand side of this equation is defined via the mean squared displacement (MSD) of the particle,
\begin{equation}
2 D^{(e)}(t) t=\left\langle (s(t)-s(0))^2\right\rangle\equiv\delta s^2(t).\label{eq:diff}
\end{equation}
$s$ denotes an arbitrary coordinate of the position of the tagged particle. $\delta s^2(t)$ denotes the 1-dimensional MSD of the 3-dimensional system in equilibrium, which is independent of direction because of isotropy.  $\langle\dots\rangle$ denotes an ensemble average, which is defined in detail below.
The mobility  on the right hand side of Eq.~\eqref{eq:Einsteineq} is probed by a constant test force $F$ which starts to act on the particle at $t=0$, and defined as the ratio of the mean velocity and the force, see also below (Eq.~\eqref{eq:def:mob}). 
For a single colloid in a solvent, the MSD grows linearly in time and the mobility is time independent, on the Brownian timescale considered here.\cite{Dhont} For higher densities of the colloids, the MSD shows the characteristic two-timescales scenario with short and long time diffusivities \cite{Dhont, zhangnaegele02} and both diffusivity and mobility are time dependent.
At even higher densities, colloidal dispersions exhibit slow cooperative dynamics and form glasses. In these, long time diffusivity and mobility vanish since the particles are trapped by the surrounding particles.\cite{Fuchs98} These metastable soft solids can easily be driven into stationary states far from equilibrium by already modest flow rates. Under shear, the system recovers ergodicity \cite{Fuchs05,Fuchs09} and the particles recover finite diffusivities \cite{Krueger09b,Kruegerprepa,Berthier02,Varnik08,Besseling07} and, as will be shown, finite mobilities for the directions perpendicular to shear. The MSD in shear direction in contrast grows cubically in time. \cite{Elrick62,Krueger09b, Krueger09c} In this paper, we will  focus on the two directions perpendicular to the shear direction. Although these show diffusive behavior, the relation \eqref{eq:Einsteineq} does not hold any more in general and much work is devoted to understanding the relation between diffusivity and mobility in non-equilibrium. The violation of the FDT is generally described by the fluctuation dissipation ratio (FDR) $X_{\rm f}(t)$. It depends on the considered observable $f$ and for the case of the Einstein relation, we define the FDR $X_{\mu}$ via
\begin{equation}
X_{\mu}(t)\frac{\partial}{\partial t}D(t)\,t={k_BT}\mu(t).\label{eq:Einstein}
\end{equation}
The equilibrium-FDT in Eq.~\eqref{eq:Einsteineq} is recovered with $X_{\mu}^{(e)}(t)\equiv1$. In non-equilibrium, $X_{\mu}(t)$ deviates from unity. This is related to the existence of non-vanishing probability currents. FDRs are hence considered a possibility to quantify the currents and the deviation  from equilibrium.\cite{Crisanti03}

The FDT-violation of colloidal glasses in a non-equilibrium steady state under shear for auto-correlation functions was discussed in previous papers.\cite{Krueger09,Krueger09c} The general finding is that at short times, the equilibrium-FDT holds with $X_{\rm f}=1$, whereas at long times, the equilibrium-FDT is violated in a special way: the FDR $\lim_{t\to\infty}X_{\rm f}(t)=\hat X_{\rm f}(\tau)=\hat X_{\rm f}$ is time-independent for the whole final relation process of the driven glass. $\tau$ is time rescaled with the timescale of the shearing. This finding is in agreement with spin-glass predictions \cite{Berthier99} as well as detailed simulations \cite{Berthier02}. While in Ref.~\citen{Berthier02}, the value of  $\hat X_{\rm f}$ was found to be independent of observable $f$, which lead to the notion of an effective temperature $T_{\rm eff}$, we indeed found a slight dependence of $\hat X_{\rm f}$ on observable.\cite{Krueger09,Krueger09c}

In the case of the Einstein relation, the MSD grows linearly in time at long times and the mobility is constant, as we will see. Due to this,  the FDR naturally approaches a constant value (different from unity) at long times. It is still interesting to see whether there is a sharp transition in the parametric plot of the two quantities from short to long time behavior as was found for the correlators \cite{Berthier02, Krueger09} and in simulations also for the Einstein relation.\cite{Berthier02} The Einstein relation for sheared glassy systems has also been studied in Ref.~\citen{Zamponi05}.

The Einstein relation is probably the FDT-example which is most easily measurable in experiments. Its non-equilibrium version has indeed been investigated for a single driven Brownian particle.\cite{Blickle07} As far as known to us, the FDT-violation for the sheared system has not been studied in experiments before. 

In this paper, we present the Einstein relation for colloidal suspensions under shear. We will therefore perform the limit of small wavevector for the relations presented previously,\cite{Krueger09,Krueger09c} without further approximation. In Sec.~\ref{sec:susc}, we give the introduction to the system under consideration and derive exact expressions for the quantities of interest. In Sec.~\ref{sec:approx}, these exact expressions will be made calculable according to previous approximations. In Sec.~\ref{sec:res}, we show and discuss our final results and we close with a summary in Sec.~\ref{sec:sum}.
\section{Microscopic starting point}\label{sec:susc}
We consider the same system as in Ref.~\citen{Krueger09} and give its description here for completeness.
It contains $N$ spherical Brownian particles of diameter $d$, dispersed in a solvent in volume $V$. The particles have bare diffusion constants $D_0={k_BT}{\mu_0}$, with mobility $\mu_0$. The interparticle force acting on particle $i$ ($i=1\dots N$) at position $\vct{r}_i$ is given by $\vct{F}_i=-\partial/\partial \vct{r}_i U(\{\vct{r}_j\})$, where $U$ is the total potential energy. We neglect hydrodynamic interactions to keep the description as simple as possible. These are also absent in the computer simulations \cite{Berthier02} to which we will compare our results. 

The external driving, viz. the shear, acts on the particles via the solvent flow velocity $v(\vct{r})=\dot\gamma y \hat{\vct{x}}$, i.e., the flow points in $x$-direction and varies in $y$-direction. $\dot\gamma$ is the shear rate. The particle distribution function $\Psi(\Gamma\equiv\{{\vct{r}_i}\},t)$ obeys the Smoluchowski equation,\cite{Dhont,Fuchs05}
\begin{eqnarray}\label{eq:smol}
\partial_t \Psi(\Gamma,t)&=&\Omega \; \Psi(\Gamma,t),\nonumber\\
\Omega&=&\Omega_e+\delta\Omega=\sum_{i}\boldsymbol{\partial}_i\cdot\left[\boldsymbol{\partial}_i-{\bf F}_i - \kap\cdot\vct{r}_i\right],\label{eq:Smolu}
\end{eqnarray}
with $\kap=\dot\gamma\hat{\vct{x}}\hat{\vct{y}}$ for the case of simple shear.
$\Omega$ is called the Smoluchowski operator and it is built up by its equilibrium part, $\Omega_e=\sum_{i}\boldsymbol{\partial}_i\cdot[\boldsymbol{\partial}_i-{\bf F}_i]$ of the system without shear and the shear term $\delta\Omega=-\sum_i\boldsymbol{\partial}_i\cdot \kap\cdot\vct{r}_i$. We introduced dimensionless units $d=k_BT=D_0=1$. There are two time-independent distributions, the equilibrium distribution $\Psi_e\propto e^{-U}$, i.e., $\Omega_e \Psi_e=0$ and the stationary distribution $\Psi_s$ with $\Omega \Psi_s=0$. Ensemble averages in equilibrium and in the stationary state are denoted
\begin{subequations}
\begin{eqnarray}
\left\langle\dots\right\rangle&=&\int d\Gamma \Psi_e(\Gamma) \dots,\\
\left\langle\dots\right\rangle^{(\dot\gamma)}&=&\int d\Gamma \Psi_s(\Gamma) \dots ,
\end{eqnarray}
\end{subequations}
respectively.
 In the stationary state, the distribution function is constant but the system is not in thermal equilibrium due to the non-vanishing probability current, which gives rise to the violation of the equilibrium-FDT.\cite{Fuchs05,Krueger09}
\subsection{Mean squared displacement}\label{sec:correlation}
While the coherent, i.e., collective dynamics of shear melted glasses has been discussed in detail,\cite{Fuchs03, Fuchs09} we focus here on the MSD of a tagged particle. Its general properties under shear for the different directions will be presented elsewhere.\cite{Krueger09b,Kruegerprepa} The MSD follows from the $q\to0$ limit of the incoherent density correlation function.\cite{Fuchs98, Voigtmann04} In the following, we consider the (1-dimensional) MSD parallel to the unit-vector $\vct{e}^i_\perp$, one of the two unit-vectors $\vct{e}^1_\perp$ and $\vct{e}^2_\perp$ spanning the plane perpendicular to the shear direction. The long time diffusivity (and also the mobility) is slightly anisotropic in this plane,\cite{Varnik08, Besseling07} but the relation between diffusivity and mobility derived below will be independent of direction. It couples the diffusivity of a certain direction to the mobility of the same direction.

From the different time dependent correlation functions \cite{Fuchs09,Kruegerprepb} after switch-on of steady shear, we distinguish different MSDs. They will enter the final formula for the stationary mobility and we introduce them briefly. In the stationary state, one measures the stationary MSD $\delta{r}_{i}^2(t)$ as the ${q}\to 0$ limit of the stationary tagged particle density correlator $C_{\bf q}(t)=\langle e^{-i\vct{q}\cdot \vct{r}_s} e^{\Om t} e^{i\vct{q}(t) \cdot \vct{r}_s}\rangle^{(\dot\gamma)}$ for the direction $\vct{q}=q\vct{e}^i_\perp$,
\begin{equation}
2D(t)t\equiv\delta r_i^2(t)= \lim_{q\to0}\frac{2-2 C_{q\vct{e}^i_\perp}(t)}{q^2}.\label{eq:MSDstat}
\end{equation}
$\vct{r}_s$ is the position of the particle and $\vct{q}(t)=\vct{q}-\vct{q}\cdot\kap t$ is the advected wavevector which enters through translational invariance of the considered infinite system.\cite{Fuchs05,Fuchs09} For the directions perpendicular to shear considered here, the wavevector is time-independent. Stationary diffusivity $D(t)$ as well as stationary mobility $\mu(t)$ should carry an index $i$ which is suppressed.  

If the MSD measurement is started a (not too large) period $t_w$ after switch-on of steady shear, one measures the two-time MSD $\delta r_i^2(t,t_w)$, where $t$ is still the correlation time, i.e., $\delta r_i^2(0,t_w)=0$. For the special case of $t_w=0$, all quantities are denoted transient. 
The transient MSD follows from the transient density correlator $C_{\bf q}(t,t_w=0)=\langle e^{-i\vct{q}\cdot \vct{r}_s} e^{\Om t} e^{i \vct{q}(t) \cdot \vct{r}_s}\rangle$ as
\begin{equation}
\delta r_i^2(t,0)= \lim_{q\to0}\frac{2-2C_{q\vct{e}^i_\perp}(t,0)}{q^2}.\label{eq:MSDtran}
\end{equation}
For finite $t_w$, we use the integration through transients (ITT) approach \cite{Fuchs05} in order to express the solution of Eq.~\eqref{eq:Smolu} a time $t_w$ after switch-on as
\begin{equation}
\Psi(t_w)=\Psi_e+\int_0^{t_w} ds e^{\Omega s} \Omega \Psi_e.\label{eq:ITT}
\end{equation}
When performing averages with $\Psi(t_w)$, one uses partial integrations to let the operators in \eqref{eq:ITT} act on what is averaged with $\Psi(t_w)$. The ITT approach has proven useful for deriving the stress under time-dependent flow.\cite{Brader07,Brader08}. The two-time MSD $\delta r_i^2(t,t_w)$ follows,
\begin{align}
&\delta r_i^2(t,t_w)=\lim_{q\to0}\frac{2-2C_{q\vct{e}^i_\perp}(t,t_w) }{q^2}\nonumber\\
&= \lim_{q\to0}\frac{2-2 \left(\left\langle e^{-iq \vct{e}^i_\perp\cdot \vct{r}_s} e^{\Om t} e^{iq \vct{e}^i_\perp\cdot \vct{r}_s}\right\rangle+\dot\gamma\int_0^{t_w}ds \left\langle\sigma_{xy} e^{\Om s} e^{-iq \vct{e}^i_\perp\cdot \vct{r}_s} e^{\Om t} e^{iq \vct{e}^i_\perp\cdot \vct{r}_s}\right\rangle\right) }{q^2}.\label{eq:MSDtwo}
\end{align}
$\sigma_{xy}=-\sum_iF_i^x y_i$ is a stress tensor element which followed from $\Omega \Psi_e=\dot\gamma \sigma_{xy}\Psi_e$. Operators act on everything to the right, except for when marked differently by bracketing. For very long waiting times, one has $\Psi(t_w\to\infty)\to\Psi_s$ in Eq.~\eqref{eq:ITT}, and $\delta r_i^2(t,\infty)=\delta r_i^2(t)$ holds. In the quiescent system, i.e., without shear,  one measures the equilibrium MSD, which follows from the equilibrium correlator $C^{(e)}_{q}(t)=\langle e^{-i\vct{q} \cdot \vct{r}_s} e^{\Om_e t} e^{i \vct{q}\cdot \vct{r}_s}\rangle$,
\begin{equation}
\delta s^2(t)= \lim_{q\to0}\frac{2-2C^{(e)}_{q}(t)}{q^2}.\label{eq:MSDeq}
\end{equation}
The un-sheared system is isotropic and only the magnitude of the wavevector enters in $C^{(e)}_{q}(t)$.
\subsection{Mobility}
Let us derive the formally exact expression for the {\it stationary} mobility $\mu(t)$ of the tagged particle in direction $\vct{e}^i_\perp$. Therefore we consider the susceptibility for tagged particle density fluctuations in this direction, $\chi_{q\vct{e}^i_\perp}(t)=\langle \frac{\partial  e^{-iq \vct{e}^i_\perp\cdot \vct{r}_s}}{\partial {\bf r}_i}\cdot\boldsymbol{\partial}_i e^{\Omega^\dagger t} e^{iq \vct{e}^i_\perp\cdot \vct{r}_s}\rangle^{(\dot\gamma)}$, as found by Agarwal in 1972,\cite{Agarwal72} see Ref.~\citen{Fuchs05} for a derivation.
The mobility $\mu(t)$ for direction $\vct{e}^i_\perp$, we are seeking here, follows from the $q\to 0$-limit of the susceptibility above,
\begin{equation}
\mu(t)=\lim_{q\to 0}\frac{\chi_{q\vct{e}^i_\perp}(t)}{q^2}.\label{eq:mobex}
\end{equation}
Physically, the mobility $\mu(t)$ is the ratio of the average velocity $\vct{v}_p= v_p\vct{e}^i_{\perp}$ of the tagged particle under the external force $\vct{F}(t)=F\vct{e}^i_{\perp}\Theta(t)$ with step function $\Theta$,  
\begin{equation}
\mu(t)=\frac{\left\langle v_p\right\rangle^{(\dot\gamma,F(t))}}{F}.\label{eq:def:mob}
\end{equation}
$\langle\dots\rangle^{(\dot\gamma,F(t))}$ denotes an average in the steady state which has been perturbed by the force $\vct{F}$.
For the sheared system, the mobility is always finite in contrast to un-sheared glasses. 

In Refs.~\citen{Krueger09,Krueger09c}, we presented the exact splitting of the susceptibility into four terms. The first represents the equilibrium-FDT and the extra term is split up into three terms. The limit of $q\to0$ can be done in a straight forward manner to yield the mobility (compare Eq.~(20) in Ref.~\citen{Krueger09c}),
\begin{align}
\mu(t)=&\frac{1}{2}\frac{\partial}{\partial t} \delta r_i^2(t)-\frac{1}{4}\left.\frac{\partial}{\partial t_w} \delta r_i^2(t,t_w)\right|_{t_w=0}-\frac{1}{4}\frac{\partial}{\partial t}\left[\delta r_i^2(t) -\delta r_i^2(t,0)\right]\nonumber\\
&-\lim\limits_{q\to0}\frac{\dot\gamma}{2q^2}\int_0^\infty ds \langle \sigma_{xy}e^{\Omega^\dagger s} (\Omega^\dagger e^{-iq \vct{e}^i_\perp\cdot \vct{r}_s})e^{\Omega^\dagger t}e^{iq \vct{e}^i_\perp\cdot \vct{r}_s}\rangle.  \label{eq:exactsus}
\end{align}
Again, the first term on the right hand side of Eq.~\eqref{eq:exactsus} is the equilibrium Einstein relation for 1-dimensional diffusion, the other three terms hence correspond to the violation of the equilibrium Einstein relation. We have identified all but one terms of the susceptibility under shear with measurable mean squared displacements. The contributions of the different terms in $\chi_{\bf q}(t)$ were additionally estimated with full microscopic mode coupling approximations in Ref.~\citen{Krueger09c}.

The last term, which has yet no clear physical meaning for finite $q$, can be connected to the force-force correlation function for $q\to0$: Performing the limit of small $q$, many terms vanish due to $\int_0^\infty ds \langle \sigma_{xy}e^{\Omega^\dagger s} 1\rangle=0$ and we are left with
\begin{align}
\mu_{4}(t)&\equiv-\lim\limits_{q\to0}\frac{\dot\gamma}{2q^2}\int_0^\infty ds \langle \sigma_{xy}e^{\Omega^\dagger s} (\Omega^\dagger e^{-iq \vct{e}^i_\perp\cdot \vct{r}_s})e^{\Omega^\dagger t}e^{iq \vct{e}^i_\perp\cdot \vct{r}_s}\rangle\nonumber\\
&=-\frac{\dot\gamma}{2}\int_0^\infty ds \langle \sigma_{xy}e^{\Omega^\dagger s} \left[\left(F^i_s e^{\Omega^\dagger t} r^i_s\right)-F^i_sr^i_s\right]\rangle\nonumber\\
&\equiv\mathcal{F}(t)-\mathcal{F}(0),
\end{align}
with $F^i_s={\bf F}_s\cdot \vct{e}^i_\perp$, $r^i_s={\bf r}_s\cdot \vct{e}^i_\perp$ and
\begin{equation}
\frac{\partial}{\partial t} \mathcal{F}(t)= -\frac{\dot\gamma}{2}\int_0^\infty ds \langle \sigma_{xy}e^{\Omega^\dagger s} F^i_s e^{\Omega^\dagger t} F^i_s\rangle.\label{eq:force}
\end{equation}
Eq.~\eqref{eq:force} follows with $\Omega^\dagger r^i_s=F^i_s$.
Thus, $\mu_{4}(t)$ can be expressed in terms of the force-force correlation function 
\begin{equation}
C_{F_s^i}(t,t_w)=\langle F^i_s e^{\Om t} F^i_s\rangle+\dot\gamma\int_0^{t_w}ds \langle\sigma_{xy} e^{\Om s} F^i_s e^{\Om t} F^i_s\rangle
\end{equation}
in the following way
\begin{align}
\mu_{4}(t)&=\mathcal{F}(t)-\mathcal{F}(0)=\int_0^t dt'\left[-\frac{\dot\gamma}{2}\int_0^\infty ds \langle \sigma_{xy}e^{\Omega^\dagger s} F^i_s e^{\Omega^\dagger t'} F^i_s\rangle\right]\nonumber\\
&=\frac{1}{2}\int_0^t dt'\left[C_{F_s^i}(t',0)-C_{F_s^i}(t',\infty)\right].\label{eq:mu4}
\end{align}
The last term in Eq.~\eqref{eq:exactsus} is hence connected to the difference of transient and stationary force-force correlation function.
For long times, the force correlation decays and $\mu_4(t)$ reaches a constant. All terms of the mobility are now connected to well defined correlation functions in an exact way. The first three terms in Eq.~\eqref{eq:exactsus} yet have a different quality compared to $\mu_{4}(t)$, which we want to stress concerning the discussion about different forms of non-equilibrium FDTs:\cite{Speck06,Speck09, Baiesi09} The mean squared displacements are much easier to be determined experimentally or in simulations than the force correlation function, which can only be found when the particle positions can be resolved accurately.
We note that the stationary force correlation function in Eq.~\eqref{eq:mu4} is equal to the one in Eq.~(13) of Ref.~\citen{Szamel}. Performing the limit of $q\to 0$ starting from Eq.~(18) of Ref.~\citen{Krueger09c}, we get directly
\begin{equation}
\mu(t)=\frac{1}{4}\frac{\partial}{\partial t} \delta r_i^2(t)-\frac{1}{2}\int_0^t dt' C_{F_s^i}(t',\infty)\label{eq:exactsus2}+\frac{1}{2}.
\end{equation}
With this, Eq.~(13) of Ref.~\citen{Szamel} is reproduced. We want to emphasize again that we judge Eq.~\eqref{eq:exactsus} more useful compared to Eq.~\eqref{eq:exactsus2}. 
\section{Approximations}\label{sec:approx}
In this section, we will derive approximations in order to find closed expressions for the exact relations above. First, we treat the mean squared displacements, and then the mobility.
\subsection{Mean squared Displacements}\label{sec:schem}
\subsubsection{Equation of motion for the transient MSD} 
Within MCT-ITT, the mode coupling approach for sheared suspensions, the general strategy is to derive equations of motion for the transient quantities. These have the advantage that they are the input for the generalized Green-Kubo relations,\cite{Fuchs05,Chong09} e.g. for the shear stress.\cite{Fuchs09} The transient quantities are also more handy to be analyzed since they show e.g. the same plateau values as the corresponding functions in equilibrium.\cite{Fuchs03,Fuchs09} In a second step, two-time and stationary quantities are derived via the ITT formula, Eq.~\eqref{eq:ITT}. The separation parameter $\varepsilon$ describes the distance from the glass transition density. It is positive for glassy and negative for fluid  states.
The derivation of the equation of motion for the coherent (i.e. collective) transient correlator $C^{coh}_{\bf q}(t,0)=\langle \rho_{\bf q}^* e^{\Om t} \rho_{\vct{q}(t)}\rangle/\langle \rho_{\bf q}^* \rho_{\bf q}\rangle$, with $\rho_{\bf q}=\sum_ie^{i\vct{q}\cdot\vct{r}_i}$ has been presented in Ref.~\citen{Fuchs09}. Ref.~\citen{Krueger09b} presents the incoherent, i.e., single particle dynamics, which will also be published in a forthcoming paper.\cite{Kruegerprepa} 
 The equation of motion for the transient mean squared displacement for direction $\vct{e}^i_\perp$ is similar to the one for the quiescent MSD\cite{Fuchs98,Goetze} and reads\cite{Krueger09b}
\begin{equation}
\delta r_i^2(t,0) + \int_0^t m_0(\dot\gamma,t-t') \delta r_i^2(t,0)=2 t. \label{eq:perp}
\end{equation}
While Eq.~\eqref{eq:perp} is still exact, one has to make approximations in order to find the memory function $m_0(\dot\gamma,t)$. The MCT-ITT route leading to its numerical evaluation is presented in Appendix A. Then, $\delta r_i^2(t,0)$ and $\delta s^2(t)$ (with $m_0(0,t)$) are derived with Eq.~\eqref{eq:perp} . $m_0(\dot\gamma,t)$ and $\tilde\sigma$ below are the two quantities needed from the Appendix, all other quantities follow from the equations presented in the main text. See Ref.~\citen{Fuchs08b} for a review on MCT-ITT. 
\subsubsection{Two-time MSD}
Having found the transient MSD, we now derive approximations in order to go from the transient to the two-time MSD. In Refs.~\citen{Krueger09, Krueger09b, Kruegerprepb}, this approximation was presented for the two-time correlator. The corresponding approximation for the MSD follows by the limit of $q\to 0$. The result, whose derivation will be presented in Ref.~\citen{Kruegerprepb}, reads 
\begin{equation}
\delta r_i^2(t,t_w)\approx\delta r_i^2(t,0)+\frac{\tilde\sigma(t_w)}{\dot\gamma}\left.\frac{\partial}{\partial t_w}\delta r_i^2(t,t_w)\right|_{t_w=0}.\label{eq:first}
\end{equation}
The factorization of $t_w$- and $t$-dependent terms yielded the pre-factor 
\begin{equation}
\tilde\sigma(t_w)=\dot\gamma\int_0^{t_w}ds\langle\sigma_{xy}e^{\Om s}\sigma_{xy}\rangle/\langle\sigma_{xy}\sigma_{xy}\rangle.\label{eq:tilde}
\end{equation}
It contains in the numerator the shear stress $\sigma(t_w)=\dot\gamma\int_0^{t_w}ds\langle\sigma_{xy}e^{\Om s}\sigma_{xy}\rangle$ measured after switch-on. It grows with $t_w$ \cite{Zausch08} and, for $t_w\to\infty$, reaches the familiar steady shear stress measured in 'flow curves' as function of shear rate. According to Eq.~\eqref{eq:first}, the two-time MSD equals the stationary one once the stress has reached its steady value.

Eq.~\eqref{eq:first} is  exact for small waiting times, where $\tilde\sigma(t_w)=\dot\gamma t_w+\mathcal{O}(t_w^2)$ holds and Eq.~\eqref{eq:first} contains the first two terms of the Taylor expansion in $t_w$. For longer waiting times, it holds also quite well.\cite{Kruegerprepb} The important connection of the waiting time derivative in Eq.~\eqref{eq:first} to easier accessible time derivatives is given by\cite{Krueger09,Krueger09c}
\begin{equation}
\left.\frac{\partial}{\partial t_w}\delta r_i^2(t,t_w)\right|_{t_w=0}\approx\frac{\partial}{\partial t}\left(\delta r_i^2(t,0)-\delta s^2(t)\right).\label{eq:waiting} 
\end{equation}
Eq.~\eqref{eq:waiting} has been successfully tested quantitatively.\cite{Kruegerprepb}  Putting Eqs.~\eqref{eq:first} and \eqref{eq:waiting} together, we finally have 
\begin{equation}
\delta r_i^2(t,t_w)\approx\delta r_i^2(t,0)+\frac{\tilde\sigma(t_w)}{\dot\gamma}\frac{\partial}{\partial t}\left(\delta r_i^2(t,0)-\delta s^2(t)\right).\label{eq:MSDstatap}
\end{equation}
In this relation, we still need to know the normalized shear stress $\tilde\sigma(t_w)$, which in MCT-ITT is expressed in terms of the coherent density correlator. Only $\tilde\sigma\equiv\tilde\sigma(t_w\to\infty)$ is needed for the graphs below, see Appendix B for its derivation. From Eqs.~\eqref{eq:perp} and \eqref{eq:MSDstatap}, the diffusive long time limit of the MSD follows,
\begin{equation}
\lim_{t\to \infty}\delta r_i^2(t,t_w)=2\hat Dt=\frac{2t}{1+m_0(\dot\gamma,z=0)}\stackrel{(\varepsilon>0, \dot\gamma\to 0)}{\propto} |\dot\gamma| t.\label{eq:scaling}
\end{equation} 
$m_0(\dot\gamma,z=0)$ is the Laplace-transform $\mathcal{L}\left\{m_0(\dot\gamma,t)\right\}(z)=\int_0^\infty dt e^{-zt} m_0(\dot\gamma,t)$ at $z=0$. It scales in the liquid with the $\alpha$-relaxation time $\tau_\alpha$ and with $|\dot\gamma|^{-1}$ in glassy states. In the glass ($\varepsilon> 0$) the scaling with shear rate in Eq.~\eqref{eq:scaling} follows. The long time limit of $\delta r_i^2(t,t_w)$ is $t_w$-independent, since the second term in Eq.~\eqref{eq:MSDstatap} approaches a constant. This must of course hold because at long times, the two-time MSD has to follow the steady diffusivity $\hat D$ for all $t_w$.
\subsection{Mobility}
We want to derive approximations for the exact Eq.~\eqref{eq:exactsus} following Refs.~\citen{Krueger09, Krueger09c} for general observables $f$. The final non-equilibrium fluctuation dissipation relation derived there compares well to the simulation results from Ref.~\citen{Berthier02}. In mode coupling estimates, it was found that the last two terms in Eq.~\eqref{eq:exactsus} have different sign and almost cancel each other.\cite{Krueger09c} We hence ignore them here as was done in Refs.~\citen{Krueger09, Krueger09c}. A key approximation is again the connection of the waiting time derivative to time derivatives, Eq.~\eqref{eq:waiting}. We finally have the approximate expression for the mobility in terms of mean squared displacements,
\begin{subequations}
\begin{eqnarray}
\mu(t)&\approx&\frac{1}{2}\frac{\partial}{\partial t} \delta r_i^2(t)-\frac{1}{4}\left.\frac{\partial}{\partial t_w}\delta r_i^2(t,t_w)\right|_{t_w=0},\\
&\approx&\frac{1}{2}\frac{\partial}{\partial t} \delta r_i^2(t)-\frac{1}{4}\frac{\partial}{\partial t}\left(\delta r_i^2(t,0)-\delta s^2(t)\right).\label{eq:ex2}
\end{eqnarray}
\end{subequations}
This is the main result of this paper.
\section{Results}\label{sec:res}
\subsection{Mobility at short and long times}
We are now ready to discuss our results.
\begin{figure}[t]
\begin{center}
\includegraphics[width=0.8\linewidth]{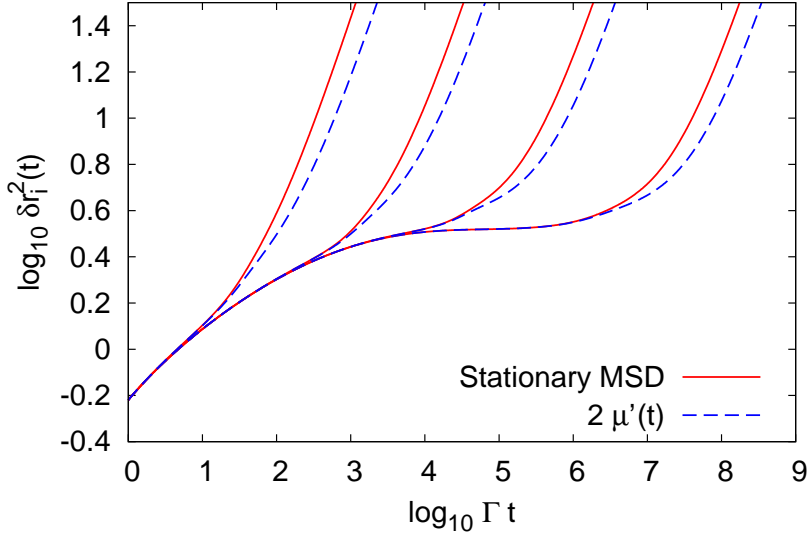}
\caption{\label{fig:FDTMSD}Stationary MSD and twice the integrated mobility $\mu'(t)=\int_0^t dt'\mu(t')$ in direction $\vct{e}^i_\perp$ for a glassy state ($\varepsilon= 10^{-3}$). Shear rates are $\dot\gamma/\Gamma=10^{-8,-6,-4,-2}$ from right to left.}
\end{center}
\end{figure} 
First, we want to illustrate the outcome of our Einstein relation Eq.~\eqref{eq:ex2} using the equations of Sec.~\ref{sec:schem} as input. The transient MSD is given as solution of Eq.~\eqref{eq:perp}, the stationary MSD follows then with Eq.~\eqref{eq:MSDstatap} and $t_w\to\infty$. The mobility is calculated with Eq.~\eqref{eq:ex2}. We present the time integrated version of Eq.~\eqref{eq:ex2} for convenience. Fig.~\ref{fig:FDTMSD} shows the stationary MSD together with the time integrated mobility for a glassy state at different shear rates. For short times, $\dot\gamma t\ll1$, we see that the equilibrium Einstein relation holds. We have $\delta r_i^2(t,0)=\delta s^2(t)+\mathcal{O}(\dot\gamma t)$\cite{Fuchs03,Krueger09b} and with Eq.~\eqref{eq:ex2},
\begin{equation}
\mu(t)=\frac{1}{2} \frac{\partial}{\partial t} \delta r_i^2(t) \hspace{1mm}\Longleftrightarrow\hspace{1mm} X_\mu(t)=1\hspace{1cm}\dot\gamma t\ll1. 
\end{equation}
For $t\agt |\dot\gamma|^{-1}$, the mobility is smaller than expected from the Einstein relation. The parametric plot in Fig.~\ref{fig:FDTMSD2} shows a rather sharp transition from short time to long time behavior with a straight line at long times corresponding to a constant FDR $X_{\mu}$. In the inset of Fig.~\ref{fig:FDTMSD2}, we see that the transition from the short to the long time value of $X_\mu$ nevertheless takes two decades in time and the strain, at which this transition happens depends strongly on shear rate. This is not apparent in the parametric plot. In Fig.~\ref{fig:FDTMSD2} we also show the mobility from Eq.~\eqref{eq:ex2} with
transient replaced by stationary MSD, from which a sharp kink in the parametric
plot and a sharper transition in the inset follows. This approximation was referred to as 'ideal $X=\frac{1}{2}$-law'
in Ref.~\citen{Krueger09c}. All our findings are reminiscent of the FDT-discussion for finite $q$. For $t\agt |\dot\gamma|^{-1}$ with $\dot\gamma\to 0$, we have $\delta s^2(t)=const.$ in Eq.~\eqref{eq:ex2} and the mobility is given by
\begin{equation}
\lim_{\dot\gamma \to 0}\mu(t)=\frac{1}{2} \frac{\partial}{\partial t} \left(\delta r_i^2(t)-\frac{1}{2} \delta r_i^2(t,0)\right) \hspace{1cm} t\agt |\dot\gamma|^{-1}.\label{eq:int}
\end{equation}
\begin{figure}[t]
\begin{center}
\includegraphics[width=0.8\linewidth]{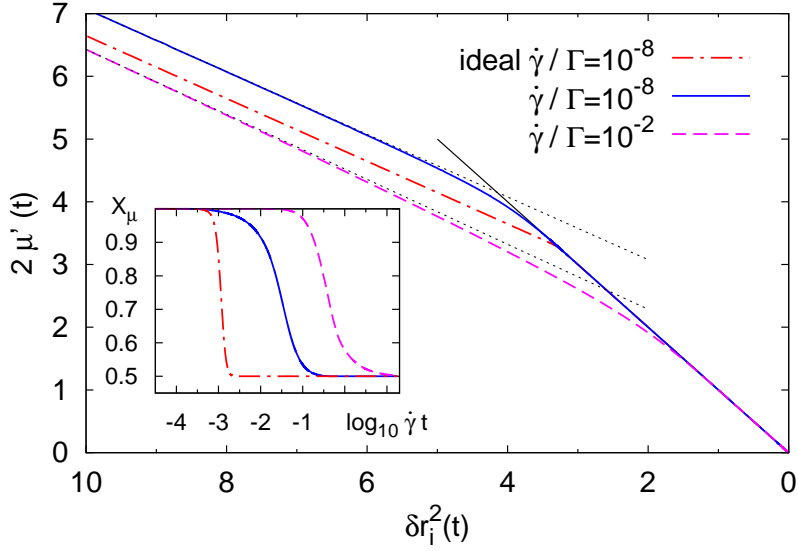}
\caption{\label{fig:FDTMSD2}Parametric plot of integrated mobility $\mu'(t)=\int_0^t dt' \mu(t')$ versus stationary MSD for a glassy state ($\varepsilon= 10^{-3}$). The dashed dotted line shows the `ideal' situation, where $\delta r_i^2(t,0)\approx \delta r_i^2(t)$ is used in Eq.~\eqref{eq:ex2}. Inset shows the FDR as function of strain $\dot\gamma t$ for the same susceptibilities.}
\end{center}
\end{figure}
For very long times, $t\gg |\dot\gamma|^{-1}$, the mobility in Eq.~\eqref{eq:int} reaches the constant $\hat\mu\propto|\dot\gamma|$ (glass). The proportionality to shear rate follows with Eq.~\eqref{eq:scaling}. The  long time Einstein relation under shear connecting $\hat D$ and $\hat\mu$ is then given by (we restore physical units only for this equation),
\begin{equation}
\hat X_\mu\hat D=k_BT \hat\mu,
\end{equation} 
with 
\begin{equation}
\hat X_{\mu}=\lim\limits_{t\to\infty}\frac{\frac{1}{2} \frac{\partial}{\partial t} \left(\delta r_i^2(t)-\frac{1}{2} \delta r_i^2(t,0)\right)}{\frac{1}{2} \frac{\partial}{\partial t} \delta r_i^2(t)}=\frac{1}{2}.\label{eq:last}
\end{equation}
The last equality followed from the equality of transient and stationary MSDs for $\dot\gamma t\gg1$. 

A comment concerning the nontrivial appearance of $\hat X_{\mu}=\frac{1}{2}$ for all glassy states in Eq.~\eqref{eq:last} is in place. Let us briefly recall the approximations which lead to Eq.~\eqref{eq:last}. First, the second term in the mobility, Eq.~\eqref{eq:exactsus}, the waiting time derivative, is expressed in terms of time derivatives, see Eq.~\eqref{eq:ex2}. This relation has been tested in switch-on simulations \cite{Kruegerprepb} and holds {\it quantitatively} for two different super-cooled liquids, at least for small shear rates. Second, the third term in Eq.~\eqref{eq:exactsus} is neglected. It is the difference between the time derivatives of transient and stationary MSDs. As stated above, at long times, the MSDs must follow $\hat D$ independent of waiting time, and the third term {\it vanishes} making this approximation exact. Third, the last term in Eq.~\eqref{eq:exactsus}, $\mu_4(t)$, is neglected. To our current knowledge, it is the only term which can give a deviation from the value of $\frac{1}{2}$ at long times, its limiting value is given with $t\to\infty$ in Eq.~\eqref{eq:mu4}. This has to be addressed in the future. For finite $q$, we found a correction to the small shear rate glass-value $\hat X=\frac{1}{2}$.\cite{Krueger09,Krueger09c} The correction followed from the difference between transient and stationary correlators, this difference is absent for the MSDs at long times. As noted before, many spin models yield $\hat X=\frac{1}{2}$ at the critical temperature.\cite{Calabrese05}
 
\begin{figure}[t]
\begin{center}
\includegraphics[width=0.8\linewidth]{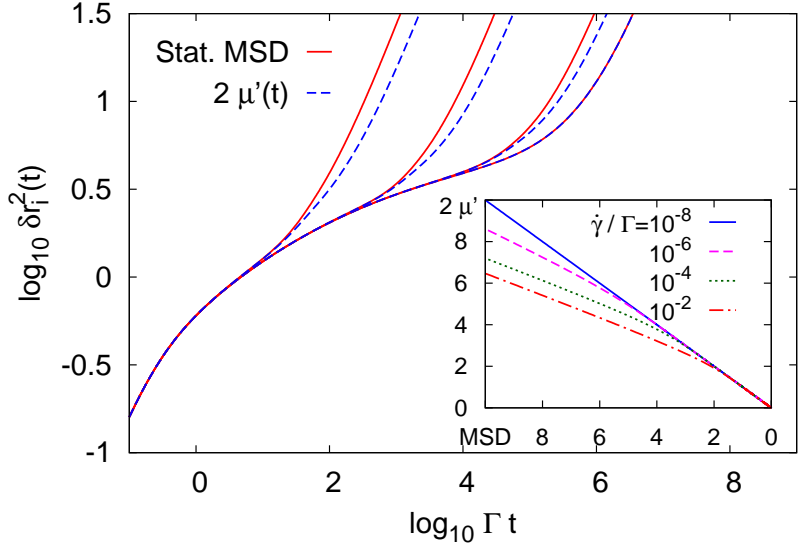}
\caption{\label{fig:FDTMSDfl}
Stationary MSD and integrated mobility $\mu'(t)=\int_0^t dt' \mu(t')$ for a fluid state ($\varepsilon=-10^{-3}$). Shear rates are $\dot\gamma/\Gamma=10^{-8,-6,-4,-2}$ from right to left. Inset shows the parametric plot for the different shear rates. For the smallest shear rate, it is almost indistinguishable from the equilibrium-FDT line.}
\end{center}
\end{figure}

Next, we consider the FDT-violation for the MSDs on the fluid side. We find the
expected behavior, i.e., for small shear rates, the equilibrium-FDT is restored for all times, see Fig.~\ref{fig:FDTMSDfl}. All our findings are in agreement with the simulation results in Ref.~\citen{Berthier02}.

\subsection{Comparison to simulations}
\begin{figure}[t]
\begin{center}
\includegraphics[width=0.8\linewidth]{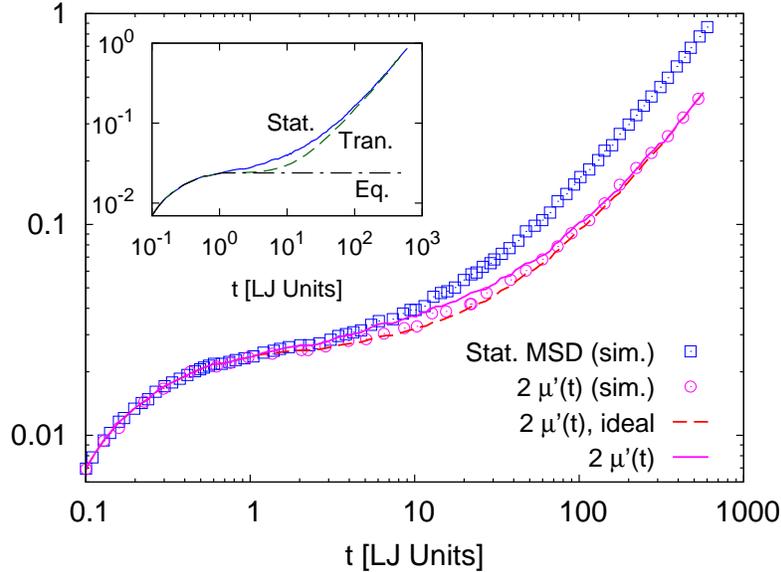}
\caption{\label{fig:FDTMSDbe}Comparison to simulation data for stationary MSD and integrated mobility $\mu'(t)=\int_0^t dt' \mu(t')$ in the neutral direction at temperature $T=0.3$ ($T_c=0.435$) and $\dot\gamma=10^{-3}$. Circles and squares are the simulation data
(including units) from Fig.~16 in Ref.~\citen{Berthier02}. The full line is the integrated mobility calculated via Eq.~(\ref{eq:ex2}). The dashed line shows the integrated mobility from Eq.~(\ref{eq:ex2}) with approximation $\delta z^2(t,0)\approx\delta z^2(t)$. Inset shows the different MSDs, see main text.}
\end{center}
\end{figure}
Following the discussion of the approximations for the different terms, it is interesting to compare Eq.~\eqref{eq:ex2} directly to the simulation results in Ref.~\citen{Berthier02}. This is done in Fig.~\ref{fig:FDTMSDbe}. We need transient as well as quiescent MSDs from the simulations as input which are not available. Therefore, we construct a quiescent MSD which is constant for long times starting on the plateau of the stationary MSD. Approximating transient and stationary MSD to be equal, we then get the dashed curve for the mobility in Fig.~\ref{fig:FDTMSDbe}, without adjustable parameter. In a second step, we can take the difference between stationary and transient MSDs into account with Eq.~\eqref{eq:MSDstatap}. We use the same value $\tilde\sigma=0.01$ as in Fig.~4 in Ref.~\citen{Krueger09}. There, we compared the susceptibility for density fluctuations at wavevector $q=7.47$ to the simulation data of Ref.~\citen{Berthier02}. Since the simulation data in Fig.~\ref{fig:FDTMSDbe} and Fig.~4 in Ref.~\citen{Krueger09} are for the same system at equal shear rate and temperature, the value of the fit-parameter $\tilde\sigma$ should be the same in both figures. The constructed quiescent and the calculated transient MSDs are shown in the inset. The resulting mobility fits very well to the simulation results. It is interesting to note that the long time mobility ($\dot\gamma t\gg1$) as calculated from \eqref{eq:ex2} is independent of our choice of the height of the non-ergodicity parameter in the quiescent MSD and of the value for $\tilde\sigma$. The agreement for long times is hence true without any fit parameter. As noted above, this agreement of the simulation data with Eq.~\eqref{eq:ex2} means that the long time value of the FDR takes the value  of $\approx\frac{1}{2}$ and that the last term in Eq.~\eqref{eq:exactsus} is indeed very small at long times. Future work has to show whether this is true in general.
\section{Summary}\label{sec:sum}
The relation between mobility and diffusivity for shear melted glasses was presented. The derived relation followed without further approximation from previous results for correlation functions by taking the small wavevector limit. At short times, the Einstein relation holds. At long times, both diffusivity and mobility are finite and scale with shear rate, but the mobility is smaller than expected from the Einstein relation. In glasses, this scaling and the violation of the Einstein relation persist to arbitrary small shear rates, i.e., the limiting value of the FDR at $\dot\gamma\to0$ jumps at the glass transition from its nontrivial value $\hat X_\mu\approx\frac{1}{2}$ to its equilibrium value $X_\mu=1$. Within MCT-ITT, this jump has the same origin as the jump of the shear stress at the glass transition: The decay of the correlators on timescale $|\dot\gamma|^{-1}$ in glasses. 

We find very good quantitative agreement to simulations. Future work has to address the only unknown term in the mobility, which is connected to the force-force correlation function.
\section*{Acknowledgments}
M.~K. was supported by the Deutsche Forschungsgemeinschaft in the International Research and Training Group 667 ``Soft Condensed Matter of Model Systems'', as well as in the Sonderforschungsbereich Transregio 6 ``Colloidal Dispersions in External Fields''. This work was partially supported by Yukawa International Program for Quark-Hadron Sciences (YIPQS).
\appendix

\section{MCT-ITT equations for density correlators}
In this appendix, the MCT-ITT equations necessary to solve Eq.~\eqref{eq:perp} are presented. In this paper, we will work on a schematic level, where the ${\bf q}$-dependence in our equations for the correlators is dropped. The ${\bf q}$-dependent treatment will be presented elsewhere.\cite{Kruegerprepa} The schematic equation of motion for the normalized coherent transient correlator $C^{coh}(t,0)\equiv\Phi^{coh}(t)$ reads \cite{Fuchs03} (this schematic model is called $F_{12}^{(\dot\gamma)}$-model)
\begin{subequations}
\label{eq:f12}
\begin{eqnarray}
0&=&\dot\Phi^{coh}(t)+\Gamma\left\{\Phi^{coh}(t)+\int_0^t dt'm^{coh}(\dot\gamma,t-t')\dot\Phi^{coh}(t')\right\},\\
m^{coh}(\dot\gamma,t)&=&\frac{1}{1+(\dot\gamma t/\gamma_c^{coh})^2}\left[(v_1^c+2.41\varepsilon)\Phi^{coh}(t)+v^c_2(\Phi^{coh}(t))^2\right],\label{eq:f122}
\end{eqnarray}
\end{subequations}
with initial decay rate $\Gamma$. We use the much studied
values $v^c_2=2$, $v^c_1=v^{c}_2(\sqrt{4/v^{c}_2}-1)$
 and take $m^{coh}(0,t)$ in order to
calculate quiescent ($\dot\gamma=0$) correlators.\cite{Goetze84} In glassy states ($\varepsilon>0$), the long time decay of $\Phi^{coh}(t)$ from the plateau down to zero happens on timescale $\dot\gamma^{-1}$,\cite{Fuchs03} i.e., without shear it stays on the plateau. For fluid states ($\varepsilon<0$), the correlator is analytic in shear rate and one observes a competition between structural relaxation on timescale $\tau_\alpha$ and shear induced relaxation on timescale $|\dot\gamma|^{-1}$. The parameter $\gamma^{coh}_c$ sets the strain $\dot\gamma t$, at which effects of shearing start to become important. We will use $\gamma^{coh}_c=1$. 

In MCT, the incoherent dynamics is coupled to the coherent one,\cite{Goetze} in the sense that the coherent correlator enters the memory function of the equation for the incoherent correlator. The equation of motion for the incoherent density correlator under shear is known,\cite{Krueger09b} the schematic version for $C(t,0)\equiv\Phi(t)$ reads\cite{Krueger09b}
\begin{subequations}
\begin{eqnarray}
0&=&\dot\Phi(t)+\Gamma\left\{\Phi(t)+\int_0^t dt'm(\dot\gamma,t-t')\dot\Phi(t')\right\},\label{eq:Sj}\\
m(\dot\gamma,t)&=&\frac{1}{1+(\dot\gamma t/\gamma_c)^2} v_s \Phi(t)\Phi^{coh}(t)\label{eq:Sj2}.
\end{eqnarray}
\end{subequations}
$m(\dot\gamma,t)$ contains the product of incoherent and coherent correlators, manifesting the coupling described above. This is why we introduce the coherent correlator here despite aiming only at tagged particle quantities. The effect of shearing is incorporated in a similar fashion as in \eqref{eq:f122} and we again use $m(0,t)$ in order to calculate quiescent quantities.\cite{Sjogren86} The additional parameter $v_s$ describes the coupling between the tagged particle and the bath particles, i.e., different size ratios can in principle be mimicked. The standard MCT-analysis for Eqs.~\eqref{eq:Sj} and \eqref{eq:Sj2}, see e.g. Ref.~\citen{Goetze}, yields the relation between the plateau values $f^{coh}$ and $f$ of $\Phi^{coh}$ and $\Phi$ respectively,\cite{Sjogren86} $f=1-\frac{1}{v_s f^{coh}}$. The incoherent dynamics is decoupled for values $v_s<v_s^c=1/f^{coh}$. For $v_1$ and $v_2$ as chosen above, we have $v^c_s\approx3$ and use $v_s=5$ which is well above $v^c_s$. For $v_s>v_s^c$, $\Phi(t)$ has similar properties as $\Phi^{coh}(t)$ for glassy and liquid states as described above.\cite{Krueger09b} Also $\gamma_c=1$. 

The memory function $m_{0}(\dot\gamma,t)$ in Eq.~\eqref{eq:perp} finally is the $q\to0$ limit of the $\bf q$-dependent analog of $m(\dot\gamma, t)$, which can be taken smoothly. We use $m_{0}(\dot\gamma,t)\equiv m(\dot\gamma, t)$ since we cannot perform the $q\to0$ limit in the schematic representation.   
\section{Normalized shear stress $\tilde\sigma$}
In the spirit of the $F_{12}^{(\dot\gamma)}$-model, we approximate the $s$-dependent normalized shear modulus by the transient correlator,\cite{Fuchs03,Krueger09}
\begin{equation}
\frac{\langle\sigma_{xy}e^{\Omega^{\dagger} s}\sigma_{xy}\rangle}{\langle\sigma_{xy}\sigma_{xy}\rangle}\approx \Phi^{coh}(s)\frac{G_\infty}{f^{coh}}\label{eq:mod},
\end{equation}
where we account for the different plateau heights of the respective normalized functions by setting $G_\infty/f^{coh}\approx\frac{1}{3}$. $\tilde\sigma$ follows then with Eq.~\eqref{eq:tilde} and $t_w\to\infty$.

\end{document}